\documentclass[12pt,a4paper]{article}

\usepackage{amsmath}
\usepackage{fancyhdr}
\usepackage{amsfonts}
 \usepackage [latin1]{inputenc}
    \usepackage{geometry}

\usepackage{amsmath}
\usepackage{amssymb}
\usepackage{makeidx}         
\usepackage{graphicx}        
\usepackage{multicol}        
\usepackage[bottom]{footmisc}
\usepackage{float}
\usepackage{tikz-cd}
\usepackage{subfig}
\usepackage{url}
\floatstyle{boxed}
\restylefloat{figure}

\makeindex             

\begin{document}

\title{The Cobb-Douglas production function revisited}
\author{Roman G. Smirnov\footnote{e-mail:  Roman.Smirnov@dal.ca}\,   and Kunpeng Wang\footnote{e-mail: kunpengwang@dal.ca} \\Department of
  Mathematics and Statistics\\
Dalhousie University\\ Halifax, Nova Scotia, Canada
  B3H~3J5}
%
%
\maketitle


\abstract{Charles Cobb and Paul Douglas in 1928 used data from the US manufacturing sector for 1899-1922 to introduce what is known today as the Cobb-Douglas production function that  has been  widely used in economic theory for decades. We employ the R programming language to fit the formulas for the parameters of the Cobb-Douglas production function generated by the authors recently via the bi-Hamiltonian approach to  the same data set utilized by Cobb and Douglas.  We conclude that the formulas for the output elastisities and total factor productivitiy are compatible with the original 1928 data.}

\section{Introduction}
\label{sec:1}

The study and applications of the Cobb-Douglas production function in  the field of economic science have a long history. Recall that in 1928 Charls Cobb and Paul Douglas published their seminal paper \cite{CD1928} in which the authors establied a relationship between  the volume of physical production in American manufacturing from 1899 to 1922 and the corresponding changes in the  amount of labor and capital that had been employed during the time period to turn out the said physical  production.  More specifically, the authors computed and expressed in logarithmic terms the index numbers  of the fixed capital,  total number of production  workers  employed  in American manufacturing,  and physical production  in manufacturing. It was established that the curve for production lied approximately one-quarter of the distance between the curves representing the corresponding changes in labor and capital. Accordingly, Cobb and Douglas adopted the  function (previously also used by Wicksteed and Wicksell) given by
\begin{equation}
\label{f1}
Y = f(L, K) = A L^k K^{1-k},
\end{equation}
where $Y$, $L$, and $K$ represented  production, labor, and capital respectively, while $A$ was total factor productivity. The authors used the method of least squares to find that  for the value of $k = 3/4$ the estimated values of $Y$ fairly well approximated the actual values for the  production in American manufacturing from 1899 to 1922. 

It took 20 more years of careful research and scrupulous study of different data before the economic community accepted the formula (\ref{f1}), although the research continued past the 1947 Douglas' presidential address  given to the American Economics Association in Chicago that marked the overall acceptance of the  results of the original research conducted in 1928  by Cobb and Douglas (see \cite{PHD1976} for a historical review and more details) and is being done in the 21st century (see, for example, Felipe and Adams \cite{FA2005}). Notably, the Cobb-Douglas aggregate production function is still being used to describe   data coming from different fields of study driven by growth in production (see, for example,  Prajneshu \cite{P2008}).

The next milestone in the development of the theory behind the Cobb-Douglas production function (\ref{f1}) that we wish to highlight in this paper is the research conducted by Ruzyo Sato \cite{RS1981} (see also Sato and Ramachandran \cite{SR2014} for more references and details) in which the author derived the Cobb-Douglas production function under the assumption of exponential growth in production, labor and capital, using some  standard teachniques from the Lie group  theory. Sato's results were further developed and  extended recently by the authors in \cite{SW2019} under the assumption of logistic rather than exponential growth in production and its factors (labor and capital). Under the asumptions specified, Sato derived  in a straightforward manner the general form of the Cobb-Douglas function. More specifically,   the function derived by Sato is of the following  form: 
\begin{equation}
\label{f2}
Y = f (L, K)  =  A L ^{\alpha} K^{\beta}, 
\end{equation} 
where $Y$, $L$ and $K$ are as before, while  $\alpha$ and $\beta$ denote the corresponding elastisities of substitution.  However, in order  to assure that the elastisities of substitution $\alpha$ and $\beta$ admitted  economically accepted values of $\alpha, \beta >0$, $\alpha + \beta = 1$ as in (\ref{f1}), Sato had to assume that the function in question was holothetic under two types of technical change simultaneously that assured  the same form for the production  function  (\ref{f2}) as in the original paper by Cobb and Douglas \cite{CD1928}. 

Recently the authors have extened  the result by Sato by employing the bi-Hamiltonian approach \cite{SW2019-2}. More specifically, it was shown that the exponential growth in  production and its factors (labor and capital) under some mild assumptions led to the same form of the Cobb-Douglas production function (\ref{f2}) without Sato's assumption of simultaneous holotheticity \cite{RS1981}.

The main goal of this paper is to establish a link between the analytic approach to the problem of the derivation of the Cobb-Douglas production function presented in \cite{SW2019-2} and the original data studied by Cobb and Douglas in \cite{CD1928} by employing the R programming language.

\section{Theoretical framework}
\label{sec:2}

In this section we briefly review the three approaches to the problem of the derivation of the Cobb-Douglas function outlined in the introduction.

 First, Cobb and Douglas in \cite{CD1928} presented a comprehensive study of the elastisity of labor and capital and how their variations affected the corresponding volume of production in American manufacturing from 1899 to 1922. In particular, they plotted the corresponding time series of production output (Day index of physical production), labor and capital on a logarithmic scale (see Chart I in \cite{CD1928}). Since we will use this data in what follows, let us first tabulate the index numbers of the industrial output in Amarican manufacturing $Y$, fixed capital $K$, and total number of manual workers $L$ on a logarithmic scale in the following table. 

\begin{table}[h]
\centering
 \begin{tabular}{||c | c | c | c||} 
 \hline
 Year & Output  $Y$ & Capital $K$ & Labour $L$ \\ [0.5ex] 
 \hline\hline
 1899 & 4.605170 & 4.605170 & 4.605170 \\ 
 \hline
 1900 & 4.615121 & 4.672829 & 4.653960 \\
 \hline
 1901 & 4.718499 & 4.736198 & 4.700480 \\
 \hline
 1902 & 4.804021 & 4.804021 & 4.770685 \\
 \hline
 1903 & 4.820282 & 4.875197 & 4.812184 \\ 
 \hline
 1904 & 4.804021 & 4.927254 & 4.753590 \\
 \hline
 1905 & 4.962845 & 5.003946 & 4.828314 \\
 \hline
 1906 & 5.023881 & 5.093750 & 4.890349\\ 
 \hline
 1907 & 5.017280 & 5.170484 & 4.927254\\
 \hline 
 1908 & 4.836282 & 5.220356 & 4.795791\\
 \hline
 1909 & 5.043425 & 5.288267 & 4.941642\\
 \hline
 1910 & 5.068904 & 5.337538 & 4.969813\\
 \hline
 1911 & 5.030438 & 5.375278 & 4.976734 \\
 \hline
 1912 & 5.176150 & 5.420535 & 5.023881 \\
 \hline 
 1913 & 5.214936 & 5.463832 & 5.036953 \\
 \hline 
 1914 & 5.129899 & 5.497168 & 5.003946 \\
 \hline 
 1915 & 5.241747 & 5.583469 & 5.036953 \\
 \hline
 1916 & 5.416100 & 5.697093 & 5.204007  \\
 \hline
 1917 & 5.424950 & 5.814131 & 5.278115 \\
 \hline
 1918 & 5.407172 & 5.902633 & 5.298317 \\
 \hline
 1919 & 5.384495 & 5.958425 & 5.262690 \\
 \hline
 1920 & 5.442418 & 6.008813 & 5.262690  \\ 
 \hline 
 1921 & 5.187386 & 6.033086 & 4.990433 \\ 
 \hline 
 1922 & 5.480639 & 6.066108 & 5.081404 \\ 
 \hline
\end{tabular}
\caption{The time series data used by Charles Cobb and Paul Douglas in \cite{CD1928}.} 
\label{table1}
\end{table}

The authors demonstrated in \cite{CD1928} with the aid of the method of least squares that the above data presented in Table \ref{table1} was subject to the following formula: 
\begin{equation}
\label{f3}
Y = f(L, K) = 1.01L^{3/4}K^{1/4},
\end{equation}
which was a special case of the formula (\ref{f2}). 

Next, recall Sato employed in \cite{RS1981}  an analytic approach to derive the Cobb-Douglas function (\ref{f2}). Summed up briefly, his approach was based on the assumption that the production and the corresponding input factors (labor and capital) grew exponentially. Under this assumption the problem of the derivation of the Cobb-Douglas function comes down to solving the following partial differential equation:

\begin{equation}
X\varphi =a K \frac{\partial \varphi}{\partial K} + b L \frac{\partial\varphi }{\partial L} + cf\frac{\partial \varphi}{\partial f} = 0,
\label{f4}
\end{equation}
where $\varphi (K, L, f) = 0$, $\partial \varphi /\partial f \not\equiv 0$ is a solution to (\ref{f4}).  Solving the corresponding system of ordinary differential equations
\begin{equation}
\frac{dK}{a K} = \frac{dL}{b L} = \frac{df}{cf},
\label{f5}
\end{equation} 
using the method of characteristics, yields the function (\ref{f2}), where $\alpha = \alpha (a,b,c), \beta = \beta(a,b,c)$. Unfortunately, the elasticity elements in this case do not attain economically meaningful values as in (\ref{f1}), because of the condition $\alpha \beta <0$.  To  mitigate  this problem Sato  introduced \cite{RS1981} the notion of the simultaneous holothenticity, which implied that a production function in question was holothetic under more than one type of technical change simultaneously. Economically, this assumption leads to  a model with the aggregate production function described by exponential, say,  growth in two different sectors of economy (or, two countries) rather than one. From the mathematical perespective, this model  yields  a production function which is  an  invariant of an integrable distribution of vector fields $\Delta$  on $\mathbb{R}^2_+$, each representing a technical change  determined by the formula  (\ref{f4}) if both of them are determined by exponential growth.  Indeed,  consider the following two vector fields, for which  a function $\varphi (K, L, f)$ is an invariant:  
\begin{equation}
X_1\varphi =K \frac{\partial \varphi}{\partial K} +  L \frac{\partial\varphi }{\partial L} + f\frac{\partial \varphi}{\partial f}=0, \quad
X_2\varphi =a K \frac{\partial \varphi}{\partial K} + b L \frac{\partial\varphi }{\partial L} + f\frac{\partial \varphi}{\partial f}=0.
\label{f6} 
\end{equation}
Clearly, the vector fields $X_1$, $X_2$ form a two-dimensional integrable distribution on $\mathbb{R}_+^2$: $[X_1, X_2] = \rho_1 X_1 + \rho_2X_2$, where $\rho_1 = \rho_2 = 0$. The corresponding total differential equation is given by  (see Chapter VII, Sato \cite{RS1981} for more details)
$$(fL - bfL)dK + (afK-fK)dL + (bKL - aKL)df = 0,$$
or, 
\begin{equation}
(1-b)\frac{dK}{K} + (a-1)\frac{dL}{L} + (b-a)\frac{df}{f} = 0. 
\label{f7}
\end{equation}
 Integrating   (\ref{f7}), we arrive at a Cobb-Douglas function of the form  (\ref{f2}), where the elasticity coefficients
$$\alpha = \frac{1-b}{a-b}, \quad \beta = \frac{a-1}{a-b}$$ satisfy the condition of constant return to scale $\alpha + \beta =1$. Of course, one has to also assume that the parameters of the exponential growth $a$ and $b$ are such that the coefficients of elastisity $\alpha, \beta >0$. 
 
Unfortunately, in spite of much ingenuity employed and a positive result, Sato's approach based on analytical methods cannot be merged with the approach by Cobb and Douglas based on a data analysis method. Indeed, the data presented in Table \ref{table1} represents  growth only in one sector of an economy and as such is incompatible with any  approach based on the notion of the simultaneous holothenticity. At the same time, it is obvious that  an additional equation must be employed to  derive the Cobb-Douglas aggregate production function with economically meaningful elastisity coefficients $\alpha$ and $\beta$ in (\ref{f2}). To resolve this contradiction, the authors of this article employed the bi-Hamiltonian approach in \cite{SW2019-2} to build on the approach introduced by Sato. 

The following is a brief review of the derivation of the Cobb-Douglas production function performed in \cite{SW2019-2}. Indeed, let us begin with Sato's  assumption about exponential growth in production, labor and capital and rewrite the PDE (\ref{f4}) as the following system of ODEs: 

\begin{equation}
\label{model1}
\dot{x}_i=b_ix_i, \quad i = 1,2,3,
\end{equation}
where $x_1 = L$ (labor), $x_2 = K$ (capital), $x_3 = f$ (production), $b_1 = b$, $b_2 = a$ and $b_3 = 1$ in Sato's notations (see (\ref{f4}). Next, we rewrite  (\ref{model1}) as the following Hamiltonian system: 
\begin{equation}
\label{model1HS}
\dot{x}_i = X^i_{H}= \pi_1^{i\ell }\frac{\partial H
}{\partial x_{\ell}}, 	\quad i = 1,2,3.
\end{equation}
Here 
\begin{equation}
\label{Poisson1}
\pi = - x_ix_j\frac{\partial}{\partial x_i}\wedge \frac{\partial} {\partial x_j}, \quad i, j = 1,2,3
\end{equation}
is the quadratic  (degenerate) Poisson bi-vector that defines the Hamiltonian function
\begin{equation}
\label{model1Hamiltonian}
H = \sum_{k=1}^3c_k\ln x_k
\end{equation}
via $X_{H} = \pi\mbox{d} H$, in which  the parameters $c_k$ are solutions to the rank 2 algebraic system $A {\bf c} = {\bf b}$ determined by  the skew-symmetric  $3 \times 3$ matrix $A$
$$
A = \begin{bmatrix} 0 & -1 & -1 \\ 
1 & 0 & - 1\\ 
1 & 1 & 0
\end{bmatrix}, 
$$ 
${\bf c} = [c_1, c_2, c_3]^{T}$ with all $c_k>0$, and ${\bf b} = [b_1, b_2, b_3]^T$, satisfying the condition 
\begin{equation}
b_1 + b_3 = b_2.
\label{c1}
\end{equation}

Alternatively, we can introduce the following new variables 
\begin{equation}
\label{Sub1}
v_i = \ln x_i, \quad i = 1, 2,3,
\end{equation}
 which lead to an even simpler form of the system (\ref{model1}), namely
\begin{equation}
\label{model1-2}
\dot{v}_i = b_i, \quad i = 1, 2, 3. 
\end{equation}
Interestingly, the substitution (\ref{Sub1}) is exactly the one used by Cobb and Douglas in \cite{CD1928}. Note that (\ref{model1-2}) is also a Hamiltonian system, provided $b_1 + b_3 = b_2$, defined by  the corresponding (degenerate) Poission bi-vector $\tilde{\pi}$ with  components $$\tilde{\pi}^{ij} = - \frac{\partial}{\partial v_i} \wedge \frac{\partial}{\partial v_j}$$ and  the corresponding  Hamiltonian $$\tilde{H} = \sum_{k=1}^3c_k v_k.$$ 

Observing that the  function  $H$ given  by (\ref{model1Hamiltonian}) is a constant of the motion of the Hamiltonian system (\ref{model1HS}), and then  solving the equation $ \sum_{k=1}^3c_k\ln x_k  = H =  \mbox{const}$ for $x_3$, we arrive at the  Cobb-Douglas production function (\ref{f2}) after the identification $x_1 = L$, $x_2 = K$, $x_3 = f$, $A = \exp\left(\frac{H_1}{c_3}\right)$, $\alpha = -\frac{c_1}{c_3}$, $\beta  = -  \frac{c_2}{c_3}$.  Next,  introduce the following bi-Hamiltonian structure for the dynamical system (\ref{model1}): 
\begin{equation}
\label{BHS1}
\dot{x}_i = X_{H_1, H_2} = \pi_1\mbox{d} H_1= \pi_2 \mbox{d}H_2, \quad i = 1, 2, 3, 
\end{equation} 
where the Hamiltonian functions $H_1$ and $H_2$  given by
\begin{equation}
\label{H1}
H_1 = b\ln x_1 + \ln x_2 + a \ln x_3, 
\quad
H_2 = \ln x_1 + a\ln x_2 + b \ln x_3.
\end{equation} 
correspond to the Poisson bi-vectors $\pi_1$ and $\pi_2$ 
\begin{equation}
\label{p1}
\pi_1 =  a_{ij} x_ix_j\frac{\partial}{\partial x_i}\wedge \frac{\partial} {\partial x_j}, \quad
\pi_2 =  b_{ij} x_ix_j\frac{\partial}{\partial x_i}\wedge \frac{\partial} {\partial x_j}, \quad i, j = 1,2,3
\end{equation}
respectively under the conditions
\begin{equation} 
\label{conditions}
\left\{\begin{array}{rcl}
bb_1 + b_2 + ab_3 & = &0, \\ b_1 + ab_2+b_3 b & = & 0. 
\end{array}\right.
\end{equation} 
Note the conditions (\ref{conditions}) (compare them to (\ref{c1})) assure that $\pi_1$ and $\pi_2$ are indeed Poisson bi-vectors compatible with the dynamics of (\ref{model1}) and corresponding to the Hamiltonians $H_1$ and $H_2$ given by (\ref{H1}).  Solving the linear system (\ref{conditions}) for $a$ and $b$ under the additional condition $b_1b_2 - b_3^2 \not=0$, we arrive at 
\begin{equation}
\label{ab}
\alpha  = \frac{a-1}{a-b} = \frac{b_3-b_1}{b_2 - b_1}, \quad \beta = \frac{1-b}{a-b} = \frac{b_3 - b_2}{b_1 - b_2}.
\end{equation}
Consider now the first integral $H_3$ given by 
\begin{equation}
H_3 = H_1 - H_2 = (b-1)\ln x_1 + (1-a)\ln x_2 + (a-b)\ln x_3. 
\label{H3}
\end{equation}
Solving the equation $H_3 = \mbox{const}$ determined by (\ref{H3}) for $x_3$, using \eqref{ab} we arrive at the Cobb-Douglas function (\ref{f2}) with the elastisities of substitution  $\alpha$ and $\beta$ given by
\begin{equation}
\label{ab1}
\alpha  = \frac{a-1}{a-b}, \quad \beta = \frac{1-b}{a-b},
\end{equation}
where $a$ and $b$ are given by (\ref{ab}). Note $\alpha + \beta = 1$,  as expected. Also, $\alpha, \beta >0$ under the additonal condition $b_2>b_3>b_1$, which implies by (\ref{model1}) that capital ($x_2 = K$) grows faster than production ($x_3 = f$), which, in turn,  grows faster than labor ($x_1 = L$).  We have also determined  the corresponding formula for total factor productivity $A$ (\ref{coeff4}). 

Our next goal is to show that the formuals obtained above via the bi-Hamiltonian approach can in fact be matched with the data employed by Cobb and Douglas in \cite{CD1928}.

\section{Main result}
\label{sec:3}

Solving the separable dynamical system  \eqref{model1}, we obtain 
\begin{equation}
\label{model2}
x_i=c_i \exp(b_i t), \quad i=1,2,3,
\end{equation}
where $c_i$ $\in$ $\mathbb{R}_+$ and $b_i$ we will determine from the data presented in Table \ref{table1}. 

Taking the logarithm (actually, much like Cobb and Douglas treated their data in \cite{CD1928})  of both sides of each equation in (\ref{model2}), we linearize them  as follows: 
\begin{equation}\label{eq4}
\ln x_i = C_i +b_i t, \quad i=1,2,3,
\end{equation}
where $C_i=\ln c_i$. 

Our next goal is to recover the corresponding values of the coefficients $C_i$, $b_i$, $i=1,2,3$ from the data presented in Table \ref{table1}. Employing R (see Appendix for more details) and the method of least squares, we arrive at the following values: 

\begin{equation}
\begin{array}{ccl}

b_1=0.02549605, & C_1=4.66953290 & \mbox{(labor)}, \\

b_2=0.06472564, &  C_2=4.61213588  & \mbox{(capital)},\\ 

b_3=0.03592651, & C_3=4.66415363 & \mbox{(production)}.
\end{array}
\end{equation}

\begin{figure}[h]
  \centering
 \includegraphics[width=\textwidth]{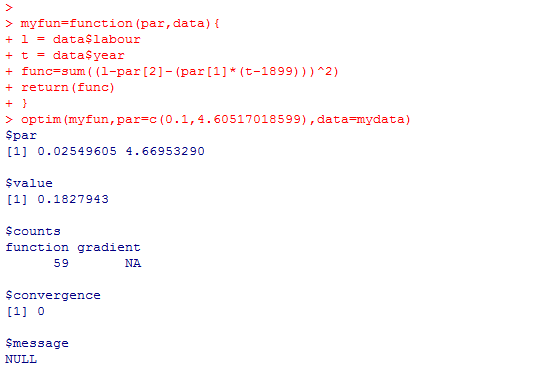}
  \caption{Labor fitting.}
  \label{figure1}
\end{figure}

\begin{figure}[h]
\centering
\includegraphics[width=\textwidth]{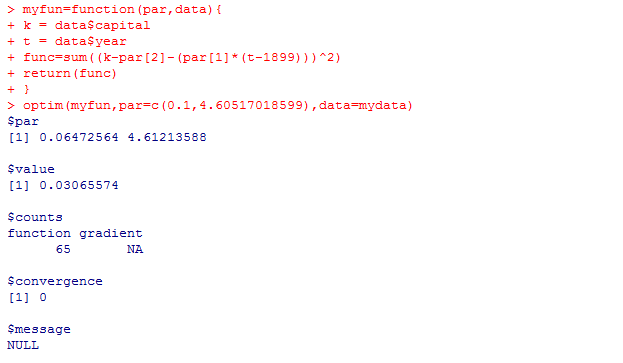}
\caption{Capital fitting.}
\label{figure2}
\end{figure}

\begin{figure}[h]
\centering
\includegraphics[width=\textwidth]{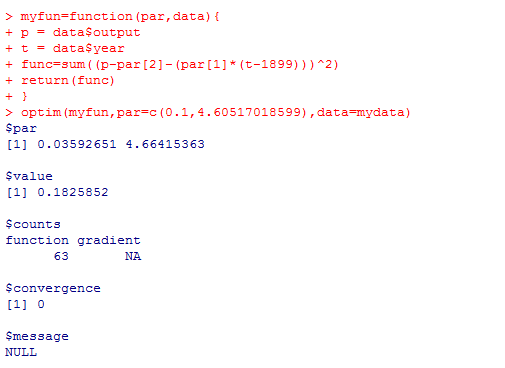}
\caption{Production fitting.}
\label{figure3}
\end{figure}

We  see that the errors, represented by the  {\$}values in Figure \ref{figure1}, \ref{figure2} and \ref{figure3},   are all  less than 1, which suggests that the formulas (\ref{eq4}) fit quite well to the data in Table \ref{table1}. To measure the goodness of fit, consider, for example,  the data presented in the second column of Table \ref{table1} (capital).  The graph relating  observed capital vs estimated capital is the subject of  Figure \ref{figure5}. Employing R, we have verified that the linear regression  shows the adjusted R-squared value of the model is 0.9934, which is very close to 1 (see Figure \ref{figure6}). 

We also note the values of the estimated coefficients satisfy the inequality $b_2>b_3>b_1$, which is in agreement with our algorithm based on the bi-Hamiltonian approach.  Identifying  $x_1=L$ and $x_2=K$ from the data and  substituting  the values of parameters $b_i$ into the equation (\ref{ab}), we obtain
\begin{equation}
a=4.659691804, \quad b=-9.104630098,
\end{equation}
which in turn determine the values of $\alpha$ and $\beta$ via (\ref{ab1}) to be
\begin{equation}
\alpha=0.2658824627, \quad \beta=0.7341175376.
\end{equation}

Now we can determine the corresponding value of total factor productivity  $A$  from the following formula, obtained by solving the equation $H_3 = \mbox{const}$ determined by (\ref{H3}), 
\begin{equation}\label{coeff4}
A=\exp\left(\frac{H_3}{a-b}\right),
\end{equation}
where $H_3$ is a constant along the flow (\ref{model1}) as a linear combination of the two Hamiltonians $H_1$ and $H_2$ given by (\ref{H1}). 

Next,  using the data  from Table \ref{table1} and  formula  (\ref{H3}), we employ R to  evaluate $H_3$, arriving at the following results: the variance of the resulting  distribution  of values of $H_3$ is $0.5923171$ and the mean of the distribution is $0.1365228$. By letting $H_3=0.1365228$ and using (\ref{coeff4}), the value of $A$ is found to be  $A=1.00996795211 \approx 1.01$ (compare with (\ref{f3})).

Therefore, we conclude  that  using statistical methods we have fitted the differential equations (\ref{model1}) to   the values of the elasticities of substitution and total factor productivity obtained via the bi-Hamiltonian approach and the data originally studied by Cobb and Douglas in 1928. In addition, we have demonstrated that Sato's assumption about exponential growth in production and factors of production \cite{RS1981} is compatible with the results by Cobb and Douglas based on the statistical analysis of the data from the US manufacturing studied in \cite{CD1928}.

\section*{Acknowledgment}
The first author (RGS) wishes to thank the organizers for the invitation to participate in the V AMMCS International Conference and present our results in the  sessions ``Applied Analysis $\&$ Inverse Problems'' and  ``Applications of Dynamical Systems $\&$ Differential Equations'', as well as to acknowledge useful discussions pertinent to this research with Professors  Herb Kunze (Guelph) and  Davide La Torre (Milan). We  also wish to thank Professor Ruzyo Sato (NYU) for his interest in our research,  comments and suggestions.

\newpage
\section*{Appendix}


\addcontentsline{toc}{section}{Appendix}

\begin{figure}[h]
\centering
\includegraphics[width=\textwidth]{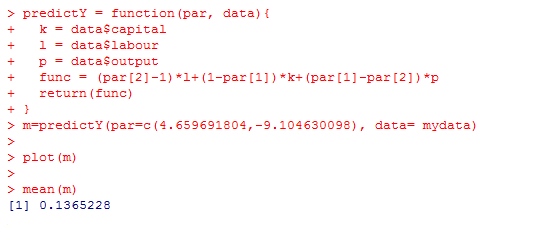}
\caption{Total factor productivity fitting.}
\label{figure4}
\end{figure}

\begin{figure}[h]
  \centering
  \caption{Observed and estimated capital versus time from 1899 to 1922.}{\label{figure5}\includegraphics[width=\textwidth]{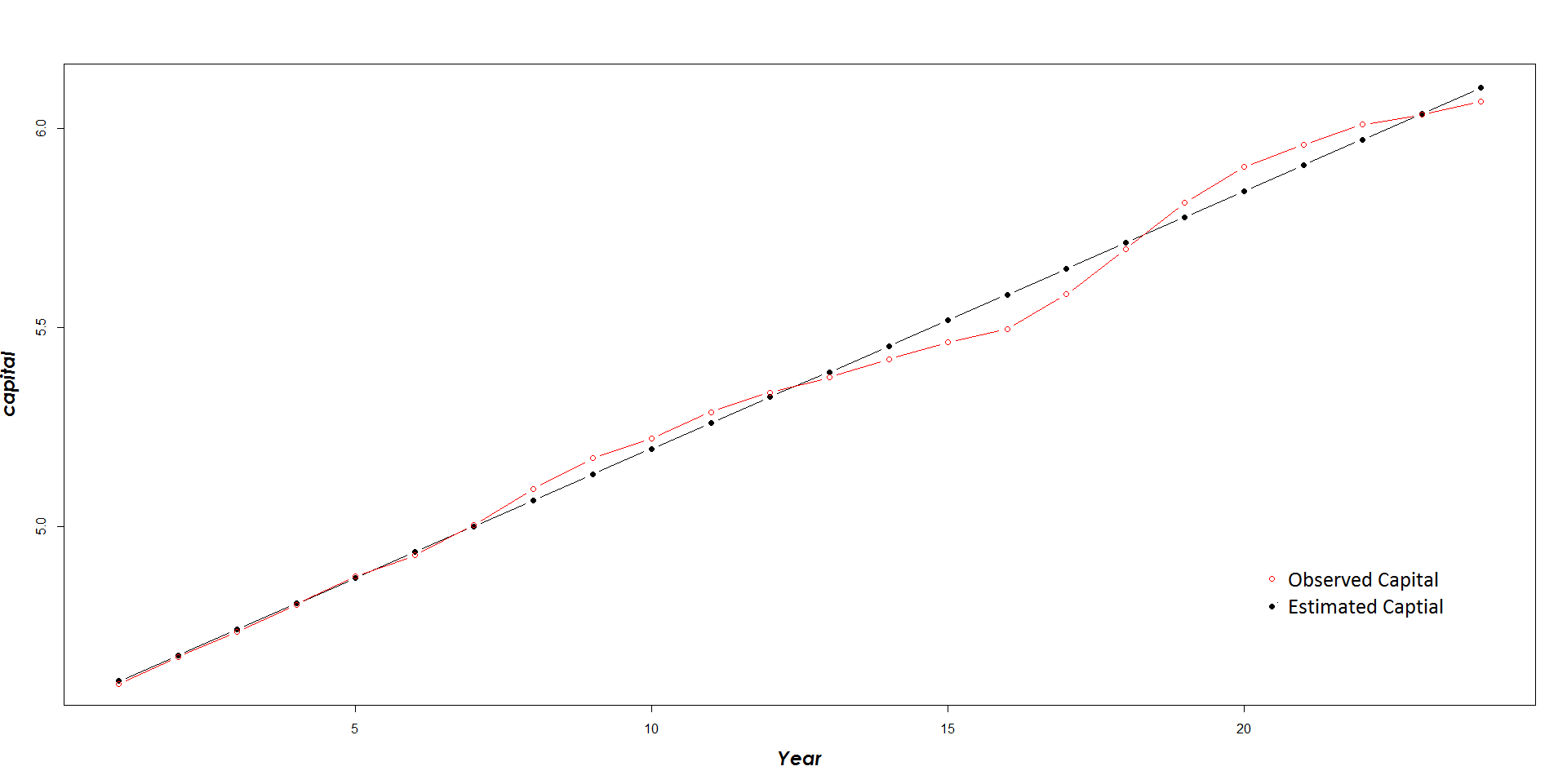}}
\end{figure}

\begin{figure}[h]
\centering
\includegraphics[width=\textwidth]{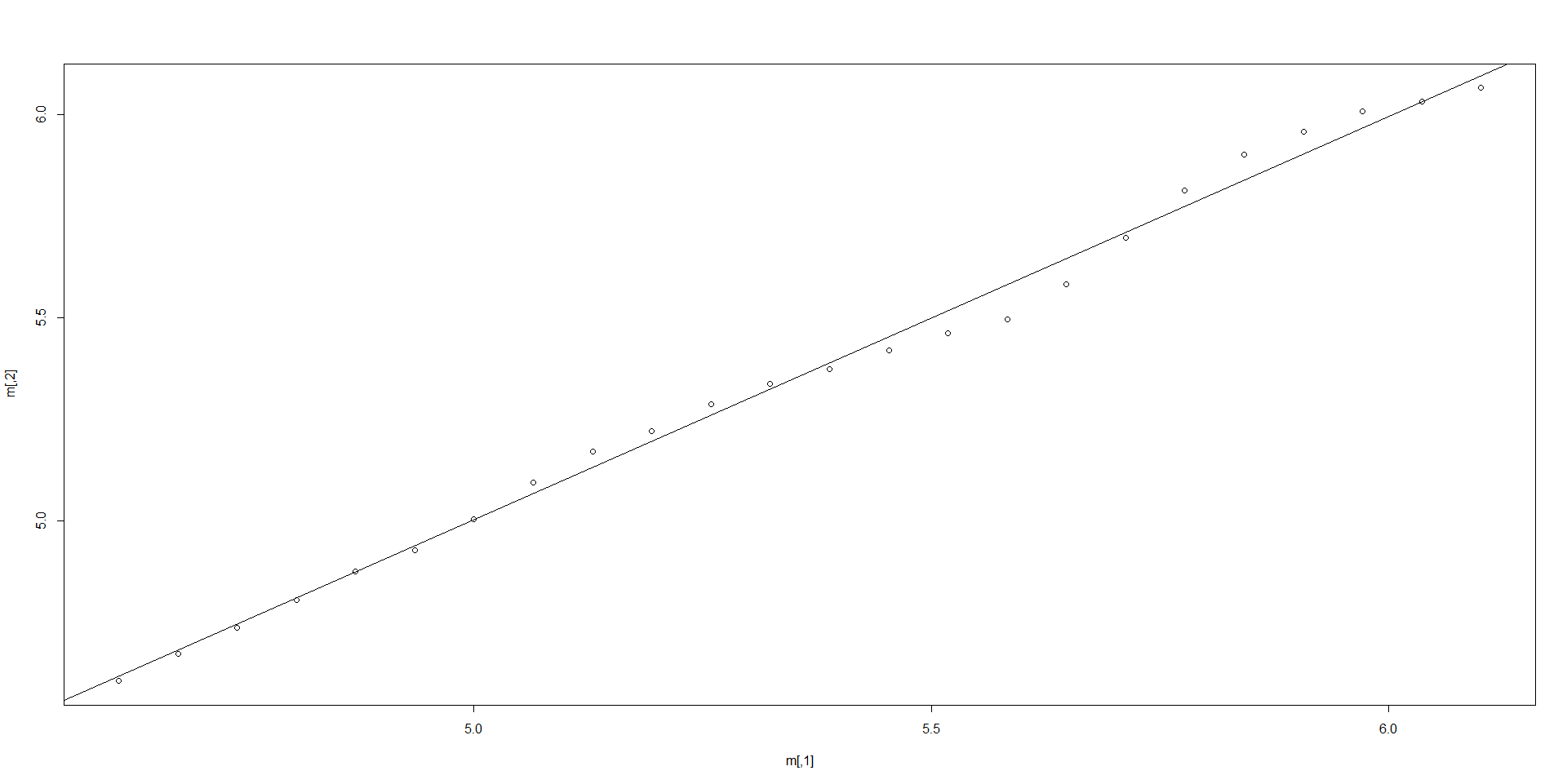}
\caption{Linear regression of the observed versus estimated capital from 1899 to 1922.}\label{figure6}
\end{figure}


\newpage
\begin{thebibliography}{99}
%
%

\bibitem{CD1928} Cobb, C. W., Douglas, P. H.: A theory of production. American Econ. Rev. \textbf{8}, 139--1965 (1928) 

\bibitem{FA2005} Felipe, J., Adams, F. G.:   The estimation of the Cobb-Douglas function: A retrospective review. Eastern Econ. J. \textbf{31}, 427--445 (2005)

\bibitem{PHD1976} Douglas, P. H.:  The Cobb-Douglas production function once again: Its history, its testing, and some new empirical values. J. Polit. Econ. \textbf{84}, 903--915 (1976)

\bibitem{P2008} Prajneshu.: Fitting of Cobb-Douglas production functions: Revisited. Agric. Econ. Res. Rev. \textbf{21}, 289--292 (2008)

\bibitem{RS1981} Sato, R.:  Theory of Technical Change and Economic Invariance. Academic Press, New York (1981)

\bibitem{SR2014} Sato, R.,   Ramachandran, R. V.: Symmetry and Economic Invariance (2nd edn). Springer, New York (2014)


\bibitem{SW2019} Smirnov, R., Wang, K.:   In search of a new economic model determined by logistic growth. European J.  Appl. Math. (2019) doi: 10.1017/S0956792519000081

\bibitem{SW2019-2} Smirnov, R., Wang K.: The Hamiltonian approach to the problem of derivation of production functions in economic growth theory. (2019)  arxiv.org/abs/1906.11224

\end{thebibliography}
\end{document}